\documentclass[journal=jctc,manuscript=article]{achemso}

\setkeys{acs}{maxauthors = 0}

\newcommand{\onlinecite}[1]{\citenum{#1}}

\usepackage[utf8]{inputenc}
\usepackage{graphicx}
\usepackage{color}
\usepackage[normalem]{ulem}
\usepackage{amsmath}
\usepackage{amssymb}

\newcommand{\R}{\mathbf{r}}
\newcommand{\funcder}[2]{\frac{\delta #1}{\delta #2}}

\newcommand{\add}[1]{\textcolor{black}{#1}}

\renewcommand{\sout}[1]{\unskip}
\newcommand{\del}[1]{\sout{#1}}
\newcommand{\ISI}{\mathrm{ISI}}
\newcommand{\SPL}{\mathrm{SPL}}

\renewcommand{\Ref}[1]{Ref. \onlinecite{#1}}

\title{Self-Consistent Implementation of Kohn-Sham Adiabatic Connection Models with Improved Treatment of the Strong-Interaction Limit}

\author{Szymon \'Smiga}
\affiliation{Institute of Physics, Faculty of Physics, Astronomy and Informatics, Nicolaus Copernicus University in Toru\'n,
ul. Grudzi\c adzka 5, 87-100 Toru\'n, Poland}
\email{szsmiga@fizyka.umk.pl}
\author{Fabio Della Sala}
\affiliation{Institute for Microelectronics and Microsystems (CNR-IMM), Via Monteroni, Campus Unisalento, 73100 Lecce, Italy}
\alsoaffiliation{Center for Biomolecular Nanotechnologies, Istituto Italiano di Tecnologia, Via Barsanti 14, 73010 Arnesano (LE), Italy}
\author{Paola Gori-Giorgi}
\affiliation{Department of Chemistry \& Pharmaceutical Sciences and Amsterdam Institute of Molecular and Life Sciences (AIMMS), Faculty of Science, Vrije Universiteit, De Boelelaan 1083, 1081HV Amsterdam, The Netherlands}
\author{Eduardo Fabiano}
\affiliation{Institute for Microelectronics and Microsystems (CNR-IMM), Via Monteroni, Campus Unisalento, 73100 Lecce, Italy}
\alsoaffiliation{Center for Biomolecular Nanotechnologies, Istituto Italiano di Tecnologia, Via Barsanti 14, 73010 Arnesano (LE), Italy}
\date{\today}

\begin{document}

\begin{abstract}
Adiabatic connection models (ACMs), which interpolate between the limits of weak and strong interaction, are powerful tools to build accurate exchange-correlation functionals. If the exact weak-interaction expansion from second-order perturbation theory is included, a self-consistent implementation of these functionals is challenging and still absent in the literature. In this work we fill this gap by presenting a fully self-consistent-field (SCF) implementation of some popular ACM functionals. While using second-order perturbation theory at weak interactions, we have also introduced new generalised gradient approximations (GGA's), beyond the usual point-charge-plus-continuum model, for the first two leading terms at strong interactions, which are crucial to ensure robustness and reliability.
We then assess the SCF-ACM functionals for molecular systems and for prototypical strong-correlation problems.
We find that they perform well for both the total energy and the electronic density
and that the impact of SCF orbitals is directly connected to the accuracy of the ACM functional form.
For the H$_2$ dissociation the SCF-ACM functionals yield significant improvements with respect to standard functionals, also thanks to the use of the new GGA's for the strong-coupling functionals.
\end{abstract}

\maketitle

\section{Introduction}
Kohn-Sham (KS)\cite{ks} density functional theory (DFT) is the most used electronic structure computational approach for molecular and solid-state systems\cite{burke12,becke14,jones15}. Its accuracy depends on the choice of the approximation for the exchange-correlation (XC) functional \cite{scuseria_xc,mardirossian_rev,mgga_rev} which, at the highest-rung of the Jacob's ladder\cite{perdew01}, involves all the occupied and virtual KS orbitals as well as the eigenvalues. Then, the XC approximation is no more an explicit functional of the density and, to stay within the pure KS formalism, the optimized effective potential (OEP) method \cite{engel:2003:3GDFT,kummel:2008:oep} must be employed.
Early OEP approaches included exact-exchange (EXX) and approximated the correlation using the second-order G\"orling-Levy perturbation theory (GL2) \cite{gl2}. However this led to a large overestimation of correlation effects and  to convergence problems \cite{grabowski:2002:OEPP2,bartlett:2005:abinit2,engel:2005:oeppt2,schweigert:2006:pt2,mori-sanchez:2005:oeppt2,grabowski:2011:jcp,grabowski:2007:ccpt2}.

Actually two different main approaches have been explored to solve this issue: 
going beyond the second-order approximation \cite{verma12,furche:2008:rpa,gruneis09,hess10,hess11,bleiz13,bleiz15,zhang16} or using a semicanonical trasformation \cite{grabowski:2002:OEPP2,bartlett:2005:abinit2,grabowski:2007:ccpt2} .
Another possible path is  the adiabatic connection (AC) formalism \cite{langreth75,gunnarsson76,savin03} which is a general, powerful tool for the development of XC functionals. \del{Since} \add{For} several decades it has been used  to justify the introduction of hybrid \cite{becke93,pbe0,fabiano15} and double hybrid (DH) functionals \cite{sharkas11,bremond11,bremond16} and successively it has been directly employed to construct high-level XC functionals based on AC models (ACM) interpolating between known limits of the AC integrand \cite{isi,isierr,seidl00,gorigiorgi09,SeiPerLev-PRA-99,liu09,ernzrehof99,teale10}.
Recently it has also been employed in the context of the Hartree-Fock (HF) theory \cite{seidl18,Daas2020} to develop corrections to the M\o ller-Plesset perturbation series \cite{spl2}.

The XC functionals based on ACMs have the general form
\begin{equation}\label{e1}
E_{xc}^{ACM} = f^{ACM}(\mathbf{W}) = \int_0^1W_\lambda^{ACM}(\mathbf{W})d\lambda\ .
\end{equation}
where $\mathbf{W}=(W_0,W'_0,W_\infty,W'_\infty)$, with
$W_0=E_x$ being the exact exchange energy,
${W'}_0=2E_c^{\rm GL2}$ being twice the GL2
correlation energy \cite{gl2}, and $W_\infty$ and ${W'}_\infty$
being the indirect part of the minimum expectation value
of the electron-electron repulsion for a given density and the potential energy of coupled zero-point oscillations around this minimum, respectively \cite{seidl07,gorigiorgi09}.
The model $W_\lambda^{ACM}$ is designed to mimic the exact but unknown $W_\lambda$, in particular by considering the known asymptotic expansions \cite{gorigiorgi09,SeiPerLev-PRA-99,gl2,seidl07}
\begin{eqnarray}
W_{\lambda\rightarrow 0} & \sim & W_0 + \lambda {W'}_0 + \cdots \\
W_{\lambda\rightarrow \infty} & \sim & W_\infty + \frac{1}{\sqrt{\lambda}}{W'}_\infty + \cdots\ .
\end{eqnarray}

In \del{the last} \add{recent} years several ACMs have been tested for various chemical applications showing promising results \cite{fabianoisi16,isigold}, especially in the description of non-covalent interactions \cite{spl2,vuckovic18}.
However, most of these recent studies have been performed within the HF-AC framework, i.e. as post-HF  calculations. Conversely, \del{only} little attention has been devoted to DFT-based ACM functionals. The main reason for this is that in the HF case the ACM is applied on top of the HF ground state \cite{seidl18,Daas2020}, which is a simple and well defined reference; on the contrary, in the DFT framework the ACM-based XC functional should be in principle applied inside the KS equations in a self-consistent-field (SCF) fashion. This requirement is not trivial because ACM-based functionals are \add{in general} not simple explicit functionals of the density but are instead complicated expressions depending on KS orbitals and orbital energies as well (through $E_x$ and $E_c^{\rm GL2}$). 
\add{One notable exception are the MCY functionals \cite{cohen07} which use semilocal approximations to set the interpolation points along the AC integrand, thus allowing for a relatively straightforward SCF implementation. In the most general case considered in this work, however, ACM functionals are fifth rung functionals} thus, in practice, also in the context of DFT, \del{ACMs} \add{they} are always applied in a post-SCF scheme using precomputed DFT densities and orbitals \cite{fabianoisi16,LUCISI}. 
In this way the results depend significantly on the choice of the reference density and orbitals, making the whole method not fully reliable \cite{fabianoisi16}.
On the other hand, an exploratory study of the XC potential derived from ACM models has shown that this possesses promising features, indicating that SCF calculations with ACM-based functionals might be an interesting path to explore \cite{potISI}.

In this work we tackle this issue by introducing an SCF implementation of the ACM potential and applying it to some test problems in order to verify its ability to describe different properties and systems.
One important aim of this work is in fact to \add{measure and } assess the \del{state-of-the-art} \add{capabilities} of some of the most popular ACM presently available 
in literature. To this purpose the use of a proper SCF procedure is crucial \add{as} \del{to really inspect} the level of accuracy of such methods \add{can be inspected} independently of an arbitrary reference ground-state as in previous works. \add{In fact, for any density functional the energy error can be decomposed into a
contribution due to the approximate nature of the functional (intrinsic error) and that due to the approximate density used in the calculation (relaxation error) \cite{kim13,sim22}. When the functional is evaluated on an arbitrary (non-SCF) density, the relaxation error may become important
and the whole performance can be influenced by the choice of the density. Indeed, recent studies have shown how this effect can be used to improve DFT results by choosing accurate non-SCF densities \cite{sim22, manni14}.
Nevertheless, within this framework its difficult to really understand the accuracy of the functional form itself and therefore to plan new advances.
On the other hand, the use of a proper SCF procedure provides a well defined reference for assessing the intrinsic errors.}
\del{and to separate the intrinsic functionals errors from the relaxation errors.} This is an extremely important point to clarify in view of future ACM developments. Note that such a development of new and possibly more accurate ACMs will instead not be covered in this work but left to upcoming publications.
The development work performed here will instead focus an a second important goal aimed at solving some open problems with the ACM potential that hinder its straightforward SCF implementation. These problems originate mainly from the naive treatment used so far for the large-$\lambda$ contributions $W_\infty$ and ${{W'}}_\infty$ which causes an unphysical behavior in the ACM potential. Hence, in this article we develop new approximations for both $W_\infty$ and ${{W'}}_\infty$ that preserve the accuracy for energies and remove the limitations on the potential side. As a byproduct of this work we obtain useful strong-correlation generalized gradient approximations that prove to be very robust for the description of the Harmonium atom and the H$_2$ dissociation.

In the following, we present the theory behind SCF implementation of ACM functionals and the construction of new $W_\infty$ and ${{W'}}_\infty$ approximations. Afterwards, we present some interesting preliminary results obtained for model and real systems.


\section{Theory}
To perform SCF ACM calculations we need to deal with the potential arising from the functional derivative of the energy of Eq. (\ref{e1}), that is \cite{potISI}
\begin{eqnarray}
\label{e5}
v_{xc,\sigma}^{\rm ACM}(\R) & \equiv & \funcder{E_{xc}^{\rm ACM}}{\rho_{\sigma}(\R)} = \\
\nonumber
& = & D^{\rm ACM}_{E_x}\funcder{E_x}{\rho_{\sigma}(\R)} + D^{\rm ACM}_{E_c^{\rm GL2}}\funcder{E_c^{\rm GL2}}{\rho_{\sigma}(\R)} +\\
\nonumber
&& + D^{\rm ACM}_{W_\infty}\funcder{W_\infty}{\rho_{\sigma}(\R)} + D^{\rm ACM}_{{W'}_\infty}\funcder{{W'}_\infty}{\rho_{\sigma}(\R)} \ ,
\end{eqnarray}
where $D_j= \partial{f^{\rm ACM}}/\partial{j}$ with $j=E_x$, $E_c^{\rm GL2}$, $W_\infty$, ${W'}_\infty$.
As discussed in Ref. \onlinecite{potISI}, the potential in Eq. (\ref{e5}) requires a combination of OEP (for $E_x$ and $E_c^{\rm GL2}$) and GGA approaches (for $W_\infty$ and ${W'}_\infty$). 
Thus it resembles the OEP-SCF implementation of the DH functionals reported in Ref. \onlinecite{DHOEPSmiga2016,DHRSOEP}.
In more details, the \del{$v_{x}(\R) = \funcder{E_x}{\rho_{\sigma}(\R)}$} \add{$v_{x\sigma}(\R) = \funcder{E_x}{\rho_{\sigma}(\R)}$}  and \del{$v_{c}(\R) =  \funcder{E_c^{\rm GL2}}{\rho_{\sigma}(\R)}$} \add{$v_{c\sigma}(\R) =  \funcder{E_c^{\rm GL2}}{\rho_{\sigma}(\R)}$} functional derivatives are obtained via solving the OEP equation which reads\cite{sharp:1953:OEP,talman:1976:OEP,grabowski:2002:OEPP2,kummel:2008:oep,lhf2,engel:2003:3GDFT}
\begin{equation}\label{e1a}
\int  X_\sigma(\R,\R')  v_{\text{A},\sigma}^{\text{OEP}}({\R'}) \mathrm{d}\R'=\Lambda_{\text{A},\sigma}(\R) \,  ,
\end{equation}
with $\text{A} = \text{X}, \text{C}$ denoting the exchange and correlation parts, respectively.
The inhomogeneity on the right hand side of Eq. (\ref{e1a}) is given by
\begin{eqnarray}\label{e2b}
\Lambda_{\text{A},\sigma}(\R)&=& \sum_{p}\bigg\{  \int \phi_{p\sigma}(\R)
 \sum_{q\ne p} \frac{\phi_{q\sigma}(\R)\phi_{q\sigma}(\R')}
{\varepsilon_{p\sigma}-\varepsilon_{q\sigma}}
\frac{\delta E_{\text{A}}}{\delta\phi_{p\sigma}(\R')} \mathrm{d}\R' \nonumber \\
&&+ \frac{\delta E_{\text{A}}}{\delta\varepsilon_{p\sigma}}|\phi_{p\sigma}(\R)|^2 \bigg \}
\end{eqnarray}
and the static KS linear response function 
\begin{equation}
X_\sigma(\R',\R) = 2\sum_{ia}\frac{\phi_{i\sigma}(\R')\phi_{a\sigma}(\R')\phi_{a\sigma}(\R)\phi_{i\sigma}(\R)}{\varepsilon_{i\sigma}-\varepsilon_{a\sigma}}\ .
\end{equation}
All quantities are evaluated using orbitals $\phi_{p\sigma}$ and eigenvalues $\varepsilon_{p\sigma}$ in a given cycle of KS SCF procedure.
(Further details can be found in Refs. \onlinecite{grabowski:2011:jcp,smiga16,grabowski14_2,DHOEPSmiga2016,DHRSOEP}).
We note, however, that there is a significant difference between ACM and DH approaches: in the former the coefficients $D_{E_x}^{ACM}$ and $D_{E_c^{GL2}}^{ACM}$ are not fixed empirical parameters as in DH, but are well defined (non-linear) functions of  $E_x,E_c^{\rm GL2},W_\infty,{W'}_\infty$ \cite{potISI}.

\subsection{Approximations for the strong-interaction limit}
Another important issue to consider in the SCF implementation of the ACMs is related to the treatment of $W_\infty$ and ${W'}_\infty$, which describe the $\lambda\to\infty$ limit of the AC integrand.
It can be proven that both $W_\infty$ and ${W'}_\infty$ display a highly non-local density dependence.\cite{ButDepGor-PRA-12,CotFriKlu-CPAM-13,CotFriKlu-ARMA-18,Lew-CRM-18,ColDiMStr-arxiv-21} This is accurately described by the strictly-correlated electrons (SCE) formalism, \cite{seidl07,gorigiorgi09} which is however computationally very demanding and non trivial to evaluate. Therefore, the $\lambda\to\infty$ limit is usually approximated by simple semilocal gradient expansions (GEA) derived within the point-charge-plus-continuum (PC) model \cite{seidl00}
%
\begin{eqnarray}
\label{e8}
W_\infty^{PC}[\rho]&=& \int d^3\R \;A \rho^{4/3}(1 + \mu_{w} s^2), \\
\label{e9}
W_\infty^{'PC}[\rho]&=& \int d^3\R \;C \rho^{3/2}(1 + \mu_{w'} s^2),
\end{eqnarray}
where  $s=|\nabla \rho|/[2(3\pi^2)^{1/3}\rho^{4/3}]$ is the reduced gradient of the density, $A=-9(4\pi/3)^{1/3}/10$, 
$C=\frac{1}{2}(3\pi)^{1/2}$,
$\mu_{w} = -3^{1/3}(2\pi)^{2/3}/35 \approx -0.1403$, and $\mu_{w'} = -0.7222$ (slightly different estimates are  possible  for $\mu_{w'}$, see  e.g. Refs. \onlinecite{isi, gorigiorgi09}).
The GEAs of Eqs. (\ref{e8}) and (\ref{e9}) yield, at least for small atoms, energies that are quite close to the accurate SCE values. However, when $s$ is large, e.g. in the tail of an exponentially decaying density, they fail, giving functional derivatives that diverge.\cite{potISI} This is a severe drawback that does not allow \del{to directly use} these approximations \add{to be used directly} in an SCF implementation. 

To remedy this limitation we consider here a simple GGA approximation, named harmonium PC (hPC) model, based on the Perdew-Burke-Ernzerhof (PBE) exchange enhancement factor \cite{pbe}, that recovers the GEAs of Eqs. (\ref{e8}) and (\ref{e9}) in the slowly-varying regime, is well behaved everywhere, and reproduces as close as possible the SCE values for both $W_{\infty}$ and ${W'}_\infty$. Thus, we have
\begin{eqnarray}
 W_\infty^{hPC}  &=& \int d^3\R\ A \rho^{4/3}\left[\frac{1+s^2 \mu_w \frac{\kappa_w+1}{\kappa_w}}{1 + s^2 \mu_w/\kappa_w }\right]  ,\\
 {W'}_\infty^{hPC}  &=&  \int d^3\R\ C \rho^{3/2}  \left[\frac{1+s^2 \mu_{w'} \frac{\kappa_{w'}+1}{\kappa_{w'}}}{1 + s^2 \mu_{w'}/\kappa_{w'} }\right]  ,
\end{eqnarray}
where $\kappa_W = -7.11$ and $\kappa_{{W'}} = -99.11$ have been fixed 
such that $W^{hPC}_{\infty}$ and $W^{'hPC}_{\infty}$ recover exactly the corresponding SCE values for the harmonium atom at $\omega = 0.5$\cite{kooi2018local}: \add{with this value the degree of correlation resembles that of the He atom and a simple analytical density is obtained.}
We note that a previous attempt to develop GGA's for $W_\infty$ and $W_\infty'$,  the modified PC (mPC) model of Ref. \onlinecite{Luc_mISI1}, yields results that are quite far from both the PC and the SCE values, in particular
$W_\infty^{'}$ does not even recover the PC model in the small $s$ limit. 
In fact, the mPC GGAs have been derived for the quasi-two-dimensional density regime \cite{Luc_mISI1}
and their application in three-dimensional systems, e.g. for the total correlation of atoms, is highly based on an error
cancellation between the quite inaccurate values of $W_{\infty}$ and ${W'}_\infty$. \cite{Luc_mISI1}
\textcolor{black}{In particular, ${W'}_\infty^{mPC}$ has been designed to compensate the inaccuracies of $W_{\infty}^{mPC}$ for the ISI functional, but this error compensation cannot work for other ACMs (especially those, as SPL, using only $W_{\infty}$).}

\begin{figure}[t]
\includegraphics[width=0.85\textwidth]{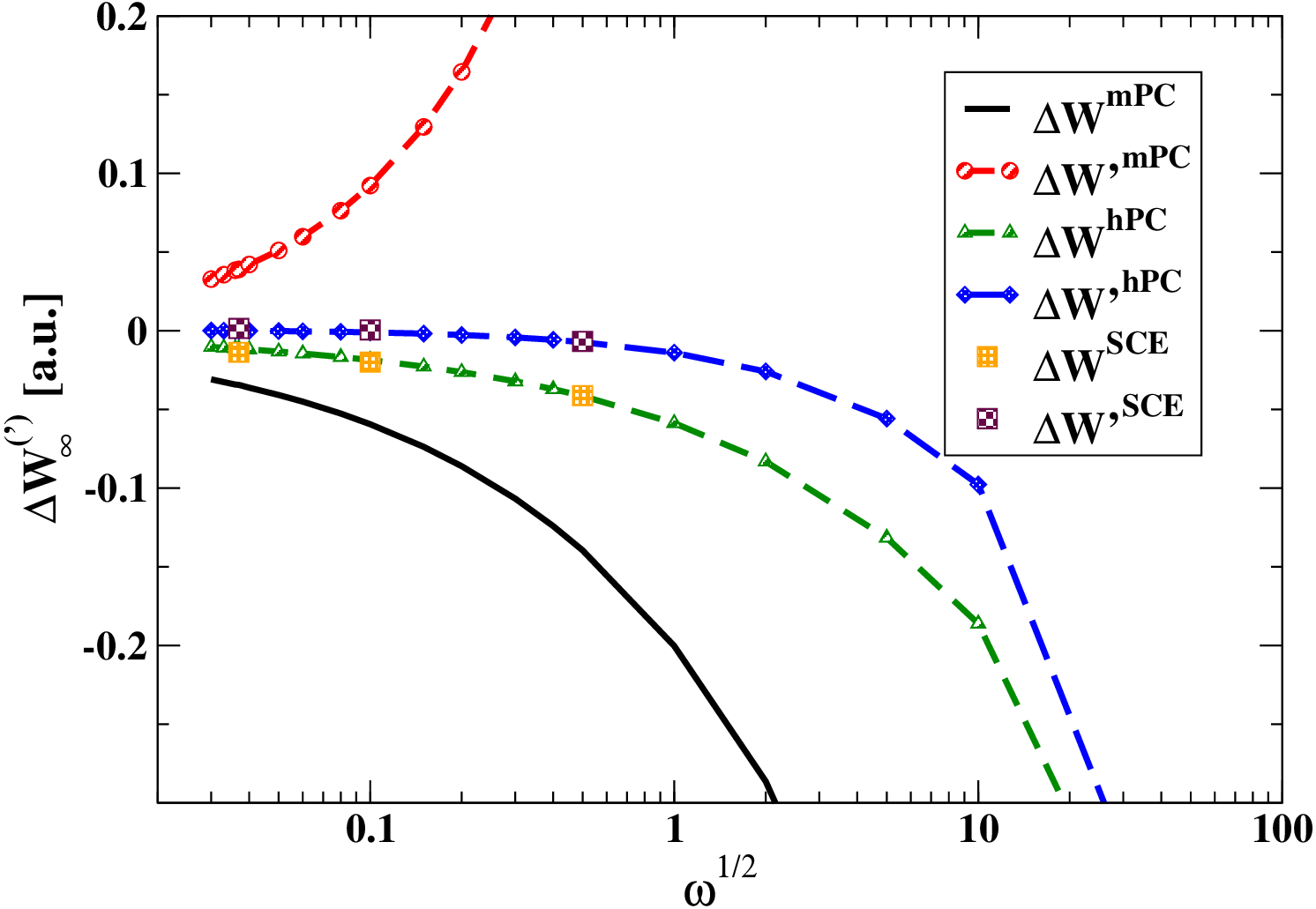}
\caption{\add{Differences between the values of $W_{\infty}$ and ${W'}_\infty$ computed with hPC and mPC formulas and the corresponding $W^{PC}_{\infty}$ and ${W'}^{PC}_{\infty}$ data ($ \Delta W^{Method}_{\infty} =  W^{Method}_{\infty} - W^{PC}_{\infty}$ ; $ \Delta {W'}^{Method}_{\infty} = {W'}^{Method}_{\infty} - {W'}^{PC}_{\infty}$) } for the harmonium atom at various values of the confinement strength $\omega$.
For reference also some available accurate SCE values are reported.\cite{kooi2018local}}
\label{fig1}
\end{figure}

To understand the performances of the different approximations for the strong-interaction functionals, we report in Fig.~\ref{fig1} the differences between the values of $W_{\infty}$ and ${W'}_\infty$ computed with the two GGA's and the PC model, for the Hooke atom at different confinement strengths $\omega$. The corresponding values for those instances of $\omega$ for which exact SCE reference data are available are also reported in Table \ref{tabs1}.
\begin{table}[hbt]
\begin{center}
\begin{tabular}{lrrrr}
\hline\hline
$\omega$ & SCE & PC & hPC & mPC \\
\hline
\multicolumn{5}{c}{$W_\infty$} \\
0.0365373 & -0.170 & -0.156 & -0.167 & -0.191  \\
0.1 & -0.304 & -0.284 & -0.303 & -0.344 \\
0.5 & -0.743 & -0.702 & -0.743 & -0.841 \\
MARE & & 6.78\% & 0.70\% & 12.90\% \\
& & & & \\
\multicolumn{5}{c}{${W'}_\infty$} \\
0.0365373 & 0.022 & 0.021 & 0.021 & 0.060 \\
0.1 & 0.054 & 0.054 & 0.053 & 0.146 \\
0.5 & 0.208 & 0.215 & 0.208 & 0.562 \\
MARE & & 2.64\% & 2.13\% & 171.10\% \\
\hline\hline
\end{tabular}
\end{center}
\caption{\label{tabs1} The $W_\infty$ \add{and} ${W'}_\infty$ energies (in Ha) for three values of $\omega$ \del{where exact} \add{for which the Hooke's atom has analytical solutions\cite{Tau-PRA-93} and exact SCE} reference data are available\cite{kooi2018local}. \add{The Hooke's atom is usually considered to be in the strong correlation regime when the density displays a maximum away from the center of the harmonic trap, which happens\cite{CioPer-JCP-00} for $\omega\lesssim 0.0401$.} The last line of each panel reports the mean absolute relative error (MARE).}
\end{table}

\begin{figure}[hbt]
\begin{center}
\includegraphics[width=0.85\textwidth]{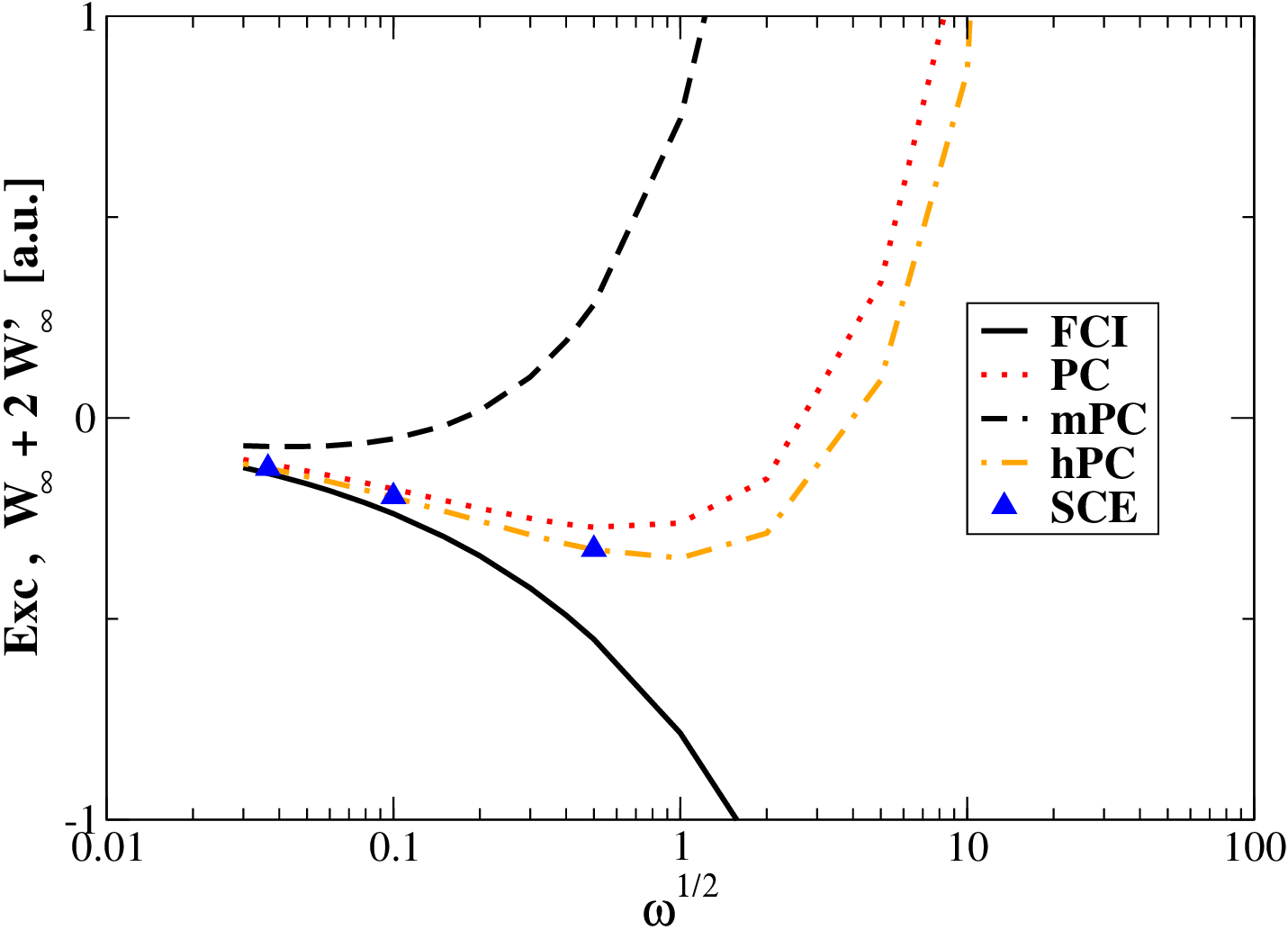}
\end{center}
\caption{\label{figs1} Comparison of the leading term of the XC energy ($E_{xc} = W_{\infty}  + 2 {W'}_{\infty}$) in the strong interacting regime of the Hooke's atom calculated using different models with FCI data\cite{LUCISI}.}
\end{figure}

\begin{figure}[hbt]
\begin{center}
\includegraphics[width=0.85\textwidth]{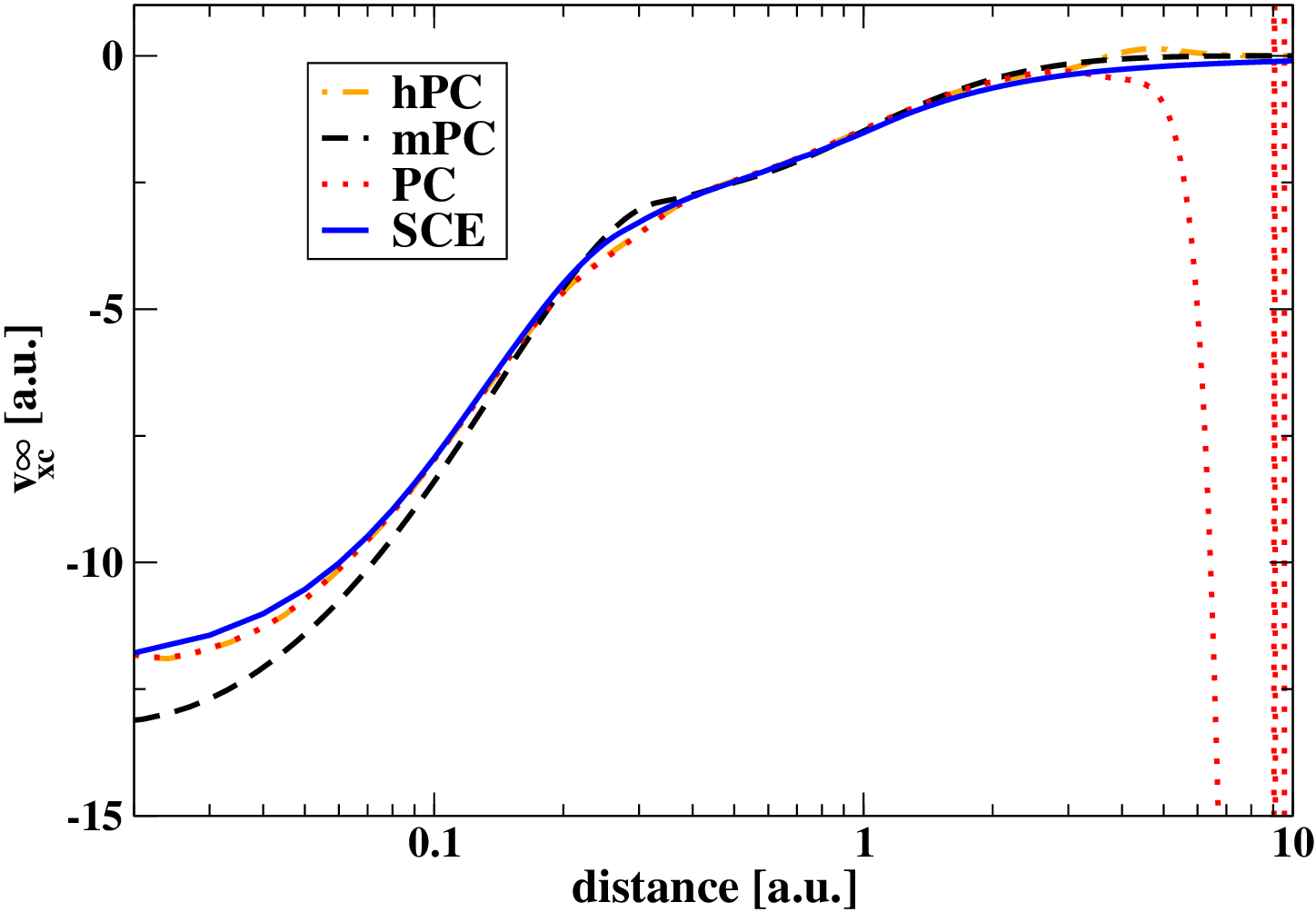}
\includegraphics[width=0.85\textwidth]{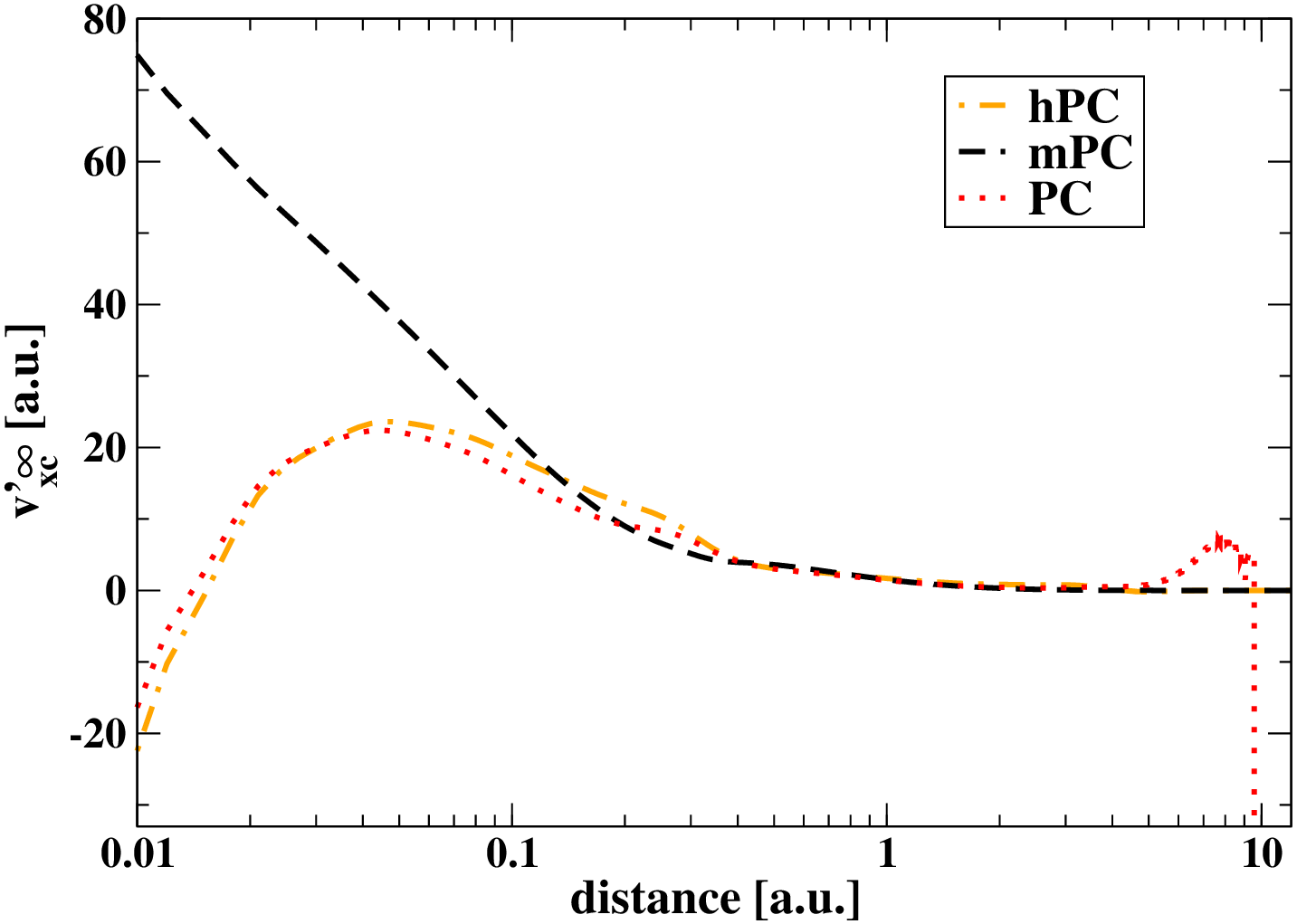}
\caption{\label{figs2} \add{Comparison between  a) $v^\infty_{xc}(\R) = \delta W_\infty / \delta \rho(\R)$ and  b) $v'^\infty_{xc}(\R)= \delta {W'}_\infty / \delta \rho(\R)$} potentials computed from different models for the Ne atom (using EXX densities).}
\end{center}
\end{figure}

We see that, unlike mPC, the hPC model \del{is much closer to the PC one and it also} reproduces very well both the $W_{\infty}$ and ${W'}_\infty$ accurate SCE values,\cite{kooi2018local} being comparable to and even superior to the original PC model. This performance is not trivial since hPC was parameterized only on a single instance of the Hooke's atom ($\omega=0.5$) but turns out to be very accurate for the whole range of confinement strengths.
In particular, Figure \ref{figs1} shows that in the small $\omega$ range (strong interaction limit of the Hooke's atom) hPC yields the best estimation of the XC energy $E_{xc} = W_{\infty}  + 2 {W'}_{\infty}$, being slightly better than PC, while the mPC method fails completely.
\add{An additional assessment is provided}
\del{Similar trends are confirmed} in Table \ref{tabs2} and Fig. \ref{figs2} \add{where real atoms are considered}
\del{also for real small atoms} both for \add{SCE} energies and \add{SCE} potentials. \del{respectively,} 
\add{Also in this case the results of the hPC functional are in line with or better than the PC model, that was originally parametrized against the He atom,}
indicating \add{once more} the robustness of the hPC method. 
\add{As anticipated the mPC is instead quite far from the reference, especially for ${{W'}}_\infty$.}
\begin{table}[hbt]
\begin{center}
\begin{tabular}{lrrrr}
\hline \hline
 & SCE & PC & hPC & mPC \\
 \hline
\multicolumn{5}{c}{$W_\infty$} \\
H & -0.3125 & \textbf{-0.3128} & -0.3293 & -0.4000 \\
He & -1.500 & -1.463 & \textbf{-1.492} & -1.671 \\
Be & \add{-4.021} & -3.943 & \textbf{-3.976} & -4.380 \\
Ne & -20.035 & \textbf{-20.018} & -20.079 & -21.022 \\
MARE & & \textbf{\add{1.15}\%} & 1.81\% & \add{13.31}\% \\
& & & & \\
\multicolumn{5}{c}{${W'}_\infty$} \\
H & 0 & 0.0426 & \textbf{0.0255} & 0.2918 \\
He & 0.621 & 0.729 & \textbf{0.646} & 1.728 \\
Be & 2.59 & 2.919 & \textbf{2.600} & 6.167 \\
Ne & 22 & 24.425 & \textbf{23.045} & 38.644 \\
MARE & & 13.71\% & \textbf{3.05\%}& 130.67\% \\
\hline\hline
\end{tabular}
\end{center}
\caption{\label{tabs2} The values of $W_{\infty}$ and ${W'}_{\infty}$ for the He, Be, and Ne atoms obtained from different models and using EXX densities. We use atomic units. The results which agree best with SCE values\cite{seidl07,gorigiorgi09} are highlighted in bold. The last line of each panel reports the mean absolute relative error (MARE) [for ${W'}_\infty$ the H results are excluded]. \add{The ${W'}^{SCE}_{\infty}$ reference data are reported with the same precision of as in the \Ref{gorigiorgi09}.}}
\end{table}

\section{Computation details}

All calculations have been performed with a locally modified ACESII \cite{acesII} software package. As in our previous studies\cite{DHOEPSmiga2016,DHRSOEP,grabowski:2011:jcp,smiga16,grabowski14_2,potISI,UHFOEP2sc,Smi-PRB-2020} in order to solve OEP equations we have employed the finite-basis set procedure of Refs. \onlinecite{gorling:1999:OEP,ivanov:1999:OEP}. In  calculations we employed the basis sets detailed below and tight convergence criteria (SCF: $10^{-8}$). In general, the \del{converges was} \add{convergence criteria were} met within several cycles of SCF procedure.

In order to solve algebraic OEP equations, the truncated singular-value decomposition (TSVD) of the response matrix was employed. The cutoff criteria in the TSVD procedure was set to $10^{-6}$. For technical details on this type of calculations we refer the reader to Refs. \onlinecite{grabowski:2011:jcp,grabowski14_2}. 

As reference data we have considered the coupled-cluster single double and perturbative triple [CCSD(T)]\cite{Raghavachari1989479} results obtained in the same basis set, in order to make a comparison on the same footing and to reduce basis set related errors. In particular, we have considered a comparison with CCSD(T) relaxed densities, the corresponding KS potentials obtained via KS inversion\cite{Wy2003} and the total CCSD(T) energies. 
In the assessment we have considered several properties i.e.:
\begin{itemize}
  \item[-]  \textbf{total energies:} the total energies have been calculated for the systems listed in Table I in Ref. \onlinecite{grabowski14_2} using identical computational setup as in the same paper.
  \add{A summary of the employed basis sets is also reported in the Supporting Information}. 
  \del{In particular, we employed an even tempered $20s10p2d$ basis for He atom and He$_2$ molecule, an uncontracted ROOS-ATZP basis~\cite{widm90} for Be, Ne atom and Ne$_2$ molecule, and an uncontracted aug-cc-pVTZ basis set~\cite{woon93} for Mg atom. For the Ar atom we used a modified basis set which combines $s$ and $p$ type basis functions from the uncontracted ROOS-ATZP~\cite{widm90} with $d$ and $f$ functions coming from the uncontracted  aug-cc-pwCVQZ basis set~\cite{peterson02}. The uncontracted cc-pVTZ basis set of Dunning~\cite{dunning:1989:bas} was used for all other systems.  For comparison, the ACM energies have been calculated at both @EXX and @SCF reference orbitals.} 
  \add{We remark that, although total energies are not very important in practical chemical applications, they are important observables and are especially useful as indicators of the quality of the ACM interpolation.}
\item[-] \add{\textbf{Dipole moments:} for selected systems (H$_2$O, HF, HCl, H$_2$S, CO) we have calculated the dipole moments using SCF densities for various methods. This is a direct test of the quality of self-consistent densities obtained within all approaches.
The uncontracted aug-cc-pVTZ basis set of Dunning~\cite{basis2} was used for all systems together with geometries taken from \Ref{NIST}.}

\item[-] \add{\textbf{HOMO and HOMO-LUMO gap energies:} as in \Ref{UPBH} and \Ref{grabowski14_2} we have computed the HOMO and HOMO-LUMO gaps, respectively, for the same set of systems as in the case of total energies. In the case of HOMO energies, the reference data have been taken from \Ref{UPBH} whereas the HOMO-LUMO gap energies have been obtained from applying the KS inversion method \cite{Wy2003} taking as a starting point the CCSD(T) relaxed density matrix as in \Ref{grabowski14_2}.}
   \item[-]  \textbf{correlation potentials and densities:} as in our previous studies\cite{grabowski:2011:jcp,grabowski14_2,grabowski:2013:molphys,Smiga2014125} , also here we investigate the quality of correlation potentials and densities\cite{Jankowski:2009:DRD,Jankowski:2010:DRD,grabowski:2011:jcp} looking at their spatial behavior. Both quantities are obtained from fully SCF
   calculations.
   The densities are analyzed in term of correlation densities defined as $\Delta \rho_c=\rho^{method}-\rho^{X}$, where $\rho^{X}$ is the density obtained from the exact exchange only (X=EXX)\cite{talman:1976:OEP} or Hartree-Fock (HF) (X=HF) calculations, for DFT and WFT methods, respectively. The Ne atom OEP calculations have been performed in fully uncontracted ROOS-ATZP \cite{widm90} \add{basis set} whereas for CO molecule the uncontracted cc-pVTZ \cite{dunning:1989:bas} basis sets was employed.
 
\item[-]  \textbf{dissociation of H$_2$:} \del{the calculation have been performed with RHF function using uncontracted aug-cc-pVTZ\cite{basis2} basis sets.}
\add{fully self-consistent and post-SCF calculations, using OEP EXX orbitals, have been performed in the spin restricted formalism using the uncontracted aug-cc-pVTZ basis set}.
For comparison PBE, MP2, GL2@EXX and FCI data are also reported.
 
  \item[-] \textbf{correlation energies of Hooke's atoms:} as previously\cite{LUCISI, PhysRevB.102.155107, Jana_2021} we have performed calculation for various values of $\omega$ in the Hooke's atom model\cite{PhysRev.128.2687} ranging between 0.03 (strong interaction) to 1000 (weak interaction) using a even-tempered Gaussian basis set from Ref. \onlinecite{B926389F}. For comparison, the ACM correlation energies have been calculated at both @EXX and @SCF reference orbitals. 

\end{itemize}

\section{Results}
We have performed a series of SCF ACM calculations to investigate the performance of these methods in the KS framework. In particular, we have considered the \del{Interaction-Strenght-Interpolarion} \add{Interaction-Strength-Interpolation} (ISI)\cite{isi} and Seidl-Perdew-Levy (SPL)\cite{SeiPerLev-PRA-99} ACMs. 
\add{Unless explicitly stated, the hPC model has been used to describe the strong-interaction limit in all calculations.}  
Moreover the bare GL2 (for SCF calculations OEP-GL2 \cite{grabowski:2002:OEPP2}) approach is also reported.
The ISI model for $W_\lambda$ has in general a larger deviation from linearity  than SPL (which does not depend on ${W'}_\infty$ too), whereas GL2  corresponds to the linear approximation $W_\lambda=2 E_{GL2}\,\lambda$. Thus the comparison of  ISI with SPL and GL2 gives information on the importance of the shape of the ACM interpolation form. 


In Table \ref{enetab} we show the total energies computed with the various methods
for a test set of 16 closed-shell atoms and small molecules, namely He, Be, Ne, Mg, Ar, HF, CO, H$_2$O, H$_2$, He$_2$,Cl$_2$,N$_2$,Ne$_2$,HCl,NH$_3$ ,C$_2$H$_6$. 
\begin{table*}
\begin{center}
\begin{small}
\begin{tabular}{lrrrcrrrr}
 \hline\hline
 & \multicolumn{3}{c}{@SCF} & $\;\;$ & \multicolumn{3}{c}{@EXX} & \\
 \cline{2-4}\cline{6-8}
System & ISI & SPL & GL2 & & ISI & SPL & GL2 & CCSD(T)\\
He & -2.90089 & -2.90043 & -2.90780 & & -2.90191 & -2.90148 & -2.90925 & -2.90253 \\
Be & -14.67318 & -14.67551 & not. conv. & & -14.67102 & -14.67278 & -14.69013 & -14.66234 \\
Ne & -128.93274 & -128.94313 & -128.98863 & & -128.92733 & -128.93628 & -128.97770 & -128.89996 \\
Mg & -199.86915 & -199.86937 & -199.88275 & & -199.86560 & -199.86569 & -199.87826 & -199.82815 \\
Ar & -527.51661 & -527.53309 & -527.58461 & & -527.51478 & -527.53095 & -527.58181 & -527.45748 \\
H$_2$ & -1.17039 & -1.16972 & -1.18107 & & -1.17019 & -1.16953 & -1.18060 & -1.17273 \\
He$_2$ & -5.80177 & -5.80086 & -5.81560 & & -5.80167 & -5.80075 & -5.81539 & -5.80506 \\
N$_2$ & -109.58263 & -109.61715 & -109.75090 & & -109.56105 & -109.58609 & -109.68725 & -109.47628 \\
Ne$_2$ & -257.86564 & -257.88644 & -257.97751 & & -257.85475 & -257.87266 & -257.95552 & -257.80003 \\
HF & -100.43787 & -100.45019 & -100.50368 & & -100.43148 & -100.44188 & -100.48965 & -100.39579 \\
CO & -113.35397 & -113.38496 & -113.51191 & & -113.32766 & -113.34760 & -113.43484 & -113.25738 \\
H$_2$O & -76.42686 & -76.44091 & -76.50285 & & -76.42076 & -76.43270 & -76.48790 & -76.38692 \\
HCl & -460.58531 & -460.58876 & -460.61411 & & -460.58227 & -460.58550 & -460.61020 & -460.50933 \\
Cl$_2$ & -919.93674 & -919.94378 & -919.99349 & & -919.92413 & -919.93022 & -919.97763 & -919.77032 \\
NH$_3$ & -56.55283 & -56.56435 & -56.62446 & & -56.54859 & -56.55876 & -56.61412 & -56.52332 \\
C$_2$H$_6$ & -79.80517 & -79.82045 & -79.92279 & & -79.79876 & -79.81239 & -79.90830 & -79.76414 \\
& & & & & & & & \\        
ME & -50.00 & -61.08$^a$ & -120.85 & & -43.14 & -52.09 & -99.17 & \\ 
MAE & 50.91 & 62.25$^a$ & 120.85 & & 43.96 & 53.16 & 99.17 & \\ 
MARE & 0.055\% & 0.071\%$^a$ & 0.162\% & & 0.048\% & 0.062\% & 0.150\% & \\ 
\hline\hline
\end{tabular} 
\end{small}   
    \caption{\label{enetab} 
    Total energies (Ha) calculated with different methods self-consistently (@SCF) or on top of EXX orbitals (@EXX), for several functionals. CCSD(T) results are given as \del{reverence} \add{reference}.  The last rows report the mean error (ME, in mHA), mean absolute error (MAE, in mHA), and the mean absolute relative error (MARE, in percent). For OEP-GL2 all the averages exclude the Be atom that for this functional has not converged.}
\end{center}
not. conv. - not converged; $^a$ without Be.
\end{table*}
We see that ISI@SCF and SPL@SCF perform quite well, \del{being twice better than} \add{giving errors roughly half that of} OEP-GL2.
\add{For comparison we acknowledge that the PBE functional \cite{pbe} yields a MARE of 0.11\%, which is twice as large as that of ISI@SCF.}
\del{This result confirms that the ACMs at the base of these functionals are quite effective to mimic the AC curve.}

Nevertheless, we have to acknowledge that the performance has further margins of improvement. For example the MAEs of MP2 and OEP2-sc (not reported) for the same test are 20 mHa and 17 mHa, respectively. 
We can trace back most of this difference to the fact that the use of KS eigenvalues, as in ISI, SPL and OEP-GL2, requires a quite large AC curvature (i.e. second derivative with respect to $\lambda$) to yield accurate results, whereas this is not the case for MP2 and OEP2-sc that employ HF-quality eigenvalues. Then, KS based methods need much more accurate ACMs to compete with HF based ones. This is also confirmed observing that in Tab. \ref{enetab}, ISI is generally better than SPL, as the former is a more advanced ACM than the latter.

A second, related observation is that the ISI and SPL results suffer from a small relaxation error that worsens slightly the performance (with respect using EXX orbitals).
\textcolor{black}{This effect might be related to the fact that the considered ACMs were developed in the context of post-SCF calculations and, as a result, may include some inherent error cancellation which is lost when they are evaluated using a (more accurate) SCF density. To understand better this trend} 
we define the quantity
\begin{equation}
\Delta[E]=    |E@{\rm SCF}-E^{{\rm ref}}| -
    |E@{\rm EXX}-E^{{\rm ref}}|
    \label{eq:deltaee}
\end{equation}
which considers the absolute error difference (with respect the reference, i.e. CCSD(T)) going from EXX orbitals to SCF orbitals (a negative value means that SCF orbitals give better accuracy than EXX orbitals).
The values of $\Delta[E]$ for ISI, SPL and OEP-GL2/GL2 are 7.0, 9.1 and 16.9 mHa, respectively. 
Despite the $\Delta[E]$ values all being positive (i.e. calculations using EXX orbitals are more accurate) they decrease going from GL2 to SPL and then from SPL to ISI, showing again that increasing the complexity/accuracy of the ACM can yield  better SCF potentials and relaxed total energies.

\begin{figure}
\begin{center}
\includegraphics[width=0.85\textwidth]{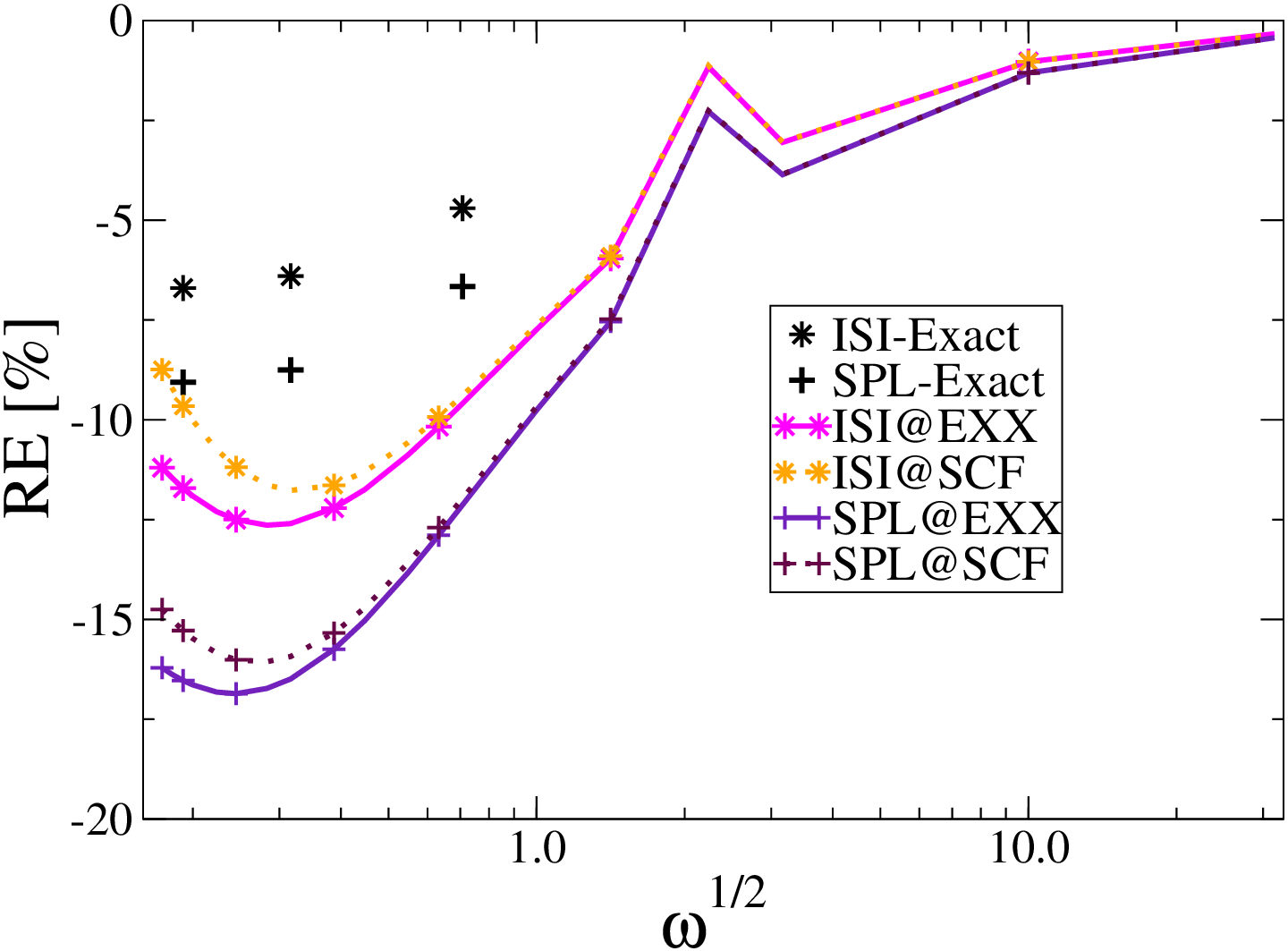}
\end{center}
\caption{\label{fighh} Relative error on correlation energies of harmonium atoms for various values of $\omega$ computed at @SCF and @EXX orbitals for ISI and SPL functionals using the hPC model for the strong-interaction functionals. \add{The errors have been computed with respect FCI data obtained in the same basis set\cite{B926389F}.} The exact ISI and SPL values are taken from Ref. \onlinecite{kooi2018local}, and are obtained by inserting exact densities into the ISI and SPL functionals, including the exact treatment (SCE) of the strong-interaction limit.}
\end{figure}

Interestingly an opposite effect of the density relaxation is found in the harmonium atom, as shown in Fig. \ref{fighh}, where 
\add{if we look at small values of the confinement strength, where the relaxation becomes more important,} 
\del{(for both ISI and SPL)} the SCF results are better with respect to the ones obtained using EXX orbitals and density \add{(for both ISI and SPL)} \del{, especially at small values of the confinement strength where the relaxation becomes more important}. 
\add{This depends on the fact that at these regimes the true density is very different from the EXX one and thus the SCF procedure produces a significant improvement on the density.}
This also traces back to the use of hPC which yields accurate strong-correlation potentials; we note in fact that the accuracy of both ACMs with the hPC model is very high (compare e.g. with Figure 3 of Ref. \onlinecite{LUCISI}). \add{Conversely using the mPC model only ISI results are rather accurate, because of error compensation effects between the $W^{mPC}_\infty$ and the ${{W'}}^{mPC}_\infty$ terms, while SPL ones, where only $W^{mPC}_\infty$ is used, are rather poor (see Fig. S2 in the Supporting Information)}.
This is an important indication of the importance of using proper strong-correlation approximations, delivering both good energies and potentials.

\add{In Table \ref{tab_dipole} we report the dipole moments of some selected systems, from the SCF density.
The results for CO are reported separately because they are qualitatively different and deserve a distinct analysis.}
\begin{table}
\begin{center}
\begin{tabular}{l|cccccc|c}
\hline\hline
Method & H$_2$O & HF & HCl & H$_2$S & \multicolumn{2}{c|}{MAE} & CO \\
\cline{6-7}
 & & & & & CCSD(T) & Exp. & \\
\hline
OEPx & 2.043 & 1.954 & 1.279 & 1.171 & 0.121 & 0.180 & -0.265 \\
GL2  & 1.616 & 1.531 & 1.061 & 1.004 & 0.187 & 0.145 & 1.703 \\
SPL & 1.758 & 1.654 & 1.085 & 1.024 & 0.110 & 0.080 & 0.940 \\
ISI & 1.809 & 1.699 & 1.093 & 1.030 & 0.083 & 0.060 & 0.692 \\
OEP2-sc & 1.885 & 1.786 & 1.185 & 1.094 & 0.018 & 0.073 & 0.355 \\
CCSD(T) & 1.904 & 1.809 & 1.170 & 1.079 &  & 0.065  & 0.153 \\
Exp. & 1.855 & 1.820 & 1.080 & 0.970 &  &  & 0.122 \\
\hline\hline
\end{tabular}
\end{center}
    \caption{Dipole moments (in Debye) for some selected systems calculated using self-consistent densities.
    Experimental data are taken from \Ref{NIST}. The mean absolute error (MAE) of H$_2$O, HF, HCl, and H$_2$S with respect to CCSD(T) and experimental results is also reported.}
    \label{tab_dipole}
\end{table}
\add{For H$_2$O, HF, HCl, and H$_2$S, a comparison with the CCSD(T) data
shows that ISI is quite effective in predicting the dipole moments being slightly better than SPL and twice as good as GL2 (for comparison PBE give in this case a MAE of 0.092 Debye with respect CCSD(T)). Anyway, as already observed for the total energies, there are important margins of improvement as testified by the OEP2-sc performance that is definitely better than the ISI one. As already discussed we can trace back the limitations of ISI and SPL in part to relaxation effects
but also on the fact that, working in a pure SCF KS framework, it is very hard for the ACM to provide a proper curvature of the AC integrand curve as to get accurate KS orbital energies; consequently the orbital-dependent energies are also negatively affected.
For the case of CO these effects are even more evident. In this case in fact OEPx predicts a qualitatively wrong dipole moment but GL2 largely over-corrects it, indicating that the linear behavior of the AC integrand needs to be significantly improved. Both ISI and SPL can partially achieve this task, halving the error with respect to GL2, but still they yield quite overestimated dipole moments.}

\begin{table*}
\begin{center}
\caption{\label{tab4} HOMO-LUMO energy gap (eV) for different systems as obtained from several methods. The last column reports the reference CCSD(T) data obtained from inverse method\cite{}.  The last lines report the mean absolute error (MAE), and the mean absolute relative error (MARE) with respect to the CCSD(T) results.}
\begin{tabular}{lccccccc}
\hline\hline
& &\multicolumn{5}{c}{@SCF} &\\
\cline{2-6} 
System & OEPx & GL2 & OEP2-sc & SPL & ISI && KS[CCSD(T)] \\
\hline
He & 21.60 & 20.95 & 21.32 & 21.23 &  21.23   && 21.21 \\ 
Be & 3.57 & not. conv. &  3.63 & 3.40 &    3.47         &&  3.61 \\ 
Ne & 18.48 & 14.12 & 16.45 & 15.17&   15.60   && 17.00 \\ 
Mg & 3.18 & 3.40 & 3.33 & 3.38 & 3.38  &&  3.36 \\ 
Ar & 11.80 & 10.95 &  11.43 & 11.08 &   11.17   &&  11.51 \\ 
H$_2$ & 12.09 & 12.03 & 12.13 & 12.12 & 12.12 &&  12.14 \\ 
He$_2$ & 21.28 &  20.64 & 21.02 & 20.81 & 20.81 &&  20.56 \\ 
N$_2$ & 9.21 & 6.73 &  8.37 & 7.68 &  7.99    &&  8.55 \\ 
Ne$_2$ & 17.84 & 13.49 & 15.75 &14.41 &  14.83&& 16.23 \\ 
HF & 11.36 & 7.80 &  9.84 & 8.70 &    9.08      &&  10.30 \\ 
CO & 7.77 & 5.87 &  7.22 & 6.68 &    6.90     &&  7.29 \\ 
H$_2$O & 8.44 & 5.99 &  7.49 & 6.73&  7.03  &&  7.75 \\ 
HCl & 7.82 & 7.10 &  7.52 & 7.11 &  7.14    && 7.55 \\ 
Cl$_2$ & 3.90 & 2.65 &  3.35 & 2.74 &  2.78  &&  3.29 \\ 
NH$_3$ & 6.97 & 5.30 &  6.35 & 5.78 & 5.98 &&  6.54 \\ 
C$_2$H$_6$ & 9.21 & 8.24 & 8.85 &8.51 & 8.62 &&  8.95 \\ 
&&&&&&&\\
ME    &+0.54   & -1.13$^a$    & -0.11 & -0.64  &  -0.48 \\
MAE &  0.52     & 1.15$^a$ &  0.21 & 0.68 & 0.52 &&   \\ 
MARE & 6.49$\%$ & 12.16\%$^a$ &  1.86\% & 7.49\% & 5.71\% &&   \\ 
\hline\hline
\end{tabular}
\end{center}
not. conv. - not converged; $^a$ without Be.
\end{table*}

\begin{table*}[htbp]
\begin{center}
\caption{\label{tab_homo} HOMO orbital energies (eV) for different systems as obtained from several approaches. In the last column we
report reference HOMO energies from Ref.\cite{smiga16}. The last lines report the mean absolute error (MAE), and the mean absolute relative error (MARE) calculated with respect to the CCSD(T) results.}
\begin{tabular}{lcccccc}
\hline\hline
&\multicolumn{5}{c}{@SCF} &\\
\cline{2-6} 
System & OEPx & GL2 & OEP2-sc & SPL & ISI & CCSD(T) \\
\hline
He & -24.98 & -24.23 & -24.55 & -24.46 & -24.39 & -24.48 \\ 
Be & -8.41 & not. conv. & -8.89 & -9.47 & -9.32 & -9.31 \\ 
Ne & -23.38 & -17.66 & -20.14 & -18.98 & -19.48 & -21.47 \\ 
Mg & -6.88 & -8.04 & -7.33 & -7.93 & -7.91 & -7.57 \\ 
Ar & -16.08 & -14.94 & -15.34 & -15.11 & -15.20 & -15.63 \\ 
H$_2$ & -16.17 & -16.34 & -16.30 & -16.25 & -16.13 & -16.41 \\ 
He$_2$ & -24.92 & -24.14 & -24.47 & -24.38 & -24.30 & -24.48 \\ 
N$_2$ & -17.17 & -11.32 & -15.65 & -13.09 & -13.78 & -15.51 \\ 
Ne$_2$ & -23.05 & -17.45 & -19.98 & -18.80 & -19.31 & -21.34 \\ 
HF & -17.48 & -12.16 & -14.57 & -13.52 & -14.03 & -15.96 \\ 
CO & -15.02 & -10.64 & -13.21 & -12.18 & -12.70 & -13.94 \\ 
H$_2$O & -13.69 & -9.01 & -11.27 & -10.39 & -10.87 & -12.50 \\ 
HCl & -12.92 & -11.94 & -12.28 & -12.04 & -12.08 & -12.59 \\ 
Cl$_2$ & -12.06 & -9.92 & -10.85 & -10.14 & -10.22 & -11.45 \\ 
NH$_3$ & -11.56 & -8.37 & -9.91 & -9.34 & -9.65 & -10.78 \\ 
C$_2$H$_6$ & -13.21 & -11.39 & -12.20 & -11.93 & -12.07 & -13.01 \\ 
 &  &  &  &  &  &  \\ 
 ME & -0.65 & +1.97$^a$ & +0.59 & +1.15 & +0.93 & \\
MAE & 0.89 & 2.49$^a$ & 0.62 & 1.22 & 0.98 &  \\ 
MARE & 6.12\% & 13.68\%$^a$ & 4.36\% & 8.28\% & 6.67\% &  \\ \hline \hline
\end{tabular}
\end{center}
not. conv. - not converged; $^a$ without Be.
\end{table*}


\add{As a next step, we consider in Table \ref{tab4} the HOMO-LUMO gaps obtained from different methods. As it could be expected both ACMs correct the general overestimation of gaps given by the OEPx but in doing so they overestimate the correlation effects yielding gaps that are too small in most cases. Thus we obtain MAEs of 0.68 and 0.52 eV for SPL and ISI, respectively, to be compared with the OEP2-sc MAE of 0.21 eV. We note anyway that the ISI and SPL results are clearly better than conventional semilocal functionals (PBE gives a MAE of 0.97 eV). 
Moreover, we note that by improving the quality of the ACM (GE2$\rightarrow$SPL$\rightarrow$ISI) the description of the HOMO-LUMO gap is also significantly improved.
Similar considerations apply as well for the HOMO energies (see Table \ref{tab_homo}). 
At the ISI level, the HOMO is shifted to higher energy with the almost the same MARE
as OEPx (which is shifted to lower energy). Again the ISI approach is better than SPL and much better than GL2 (as well as PBE with a MARE of 38.3\%.) }

\begin{figure}[t]
\includegraphics[width=0.85\textwidth]{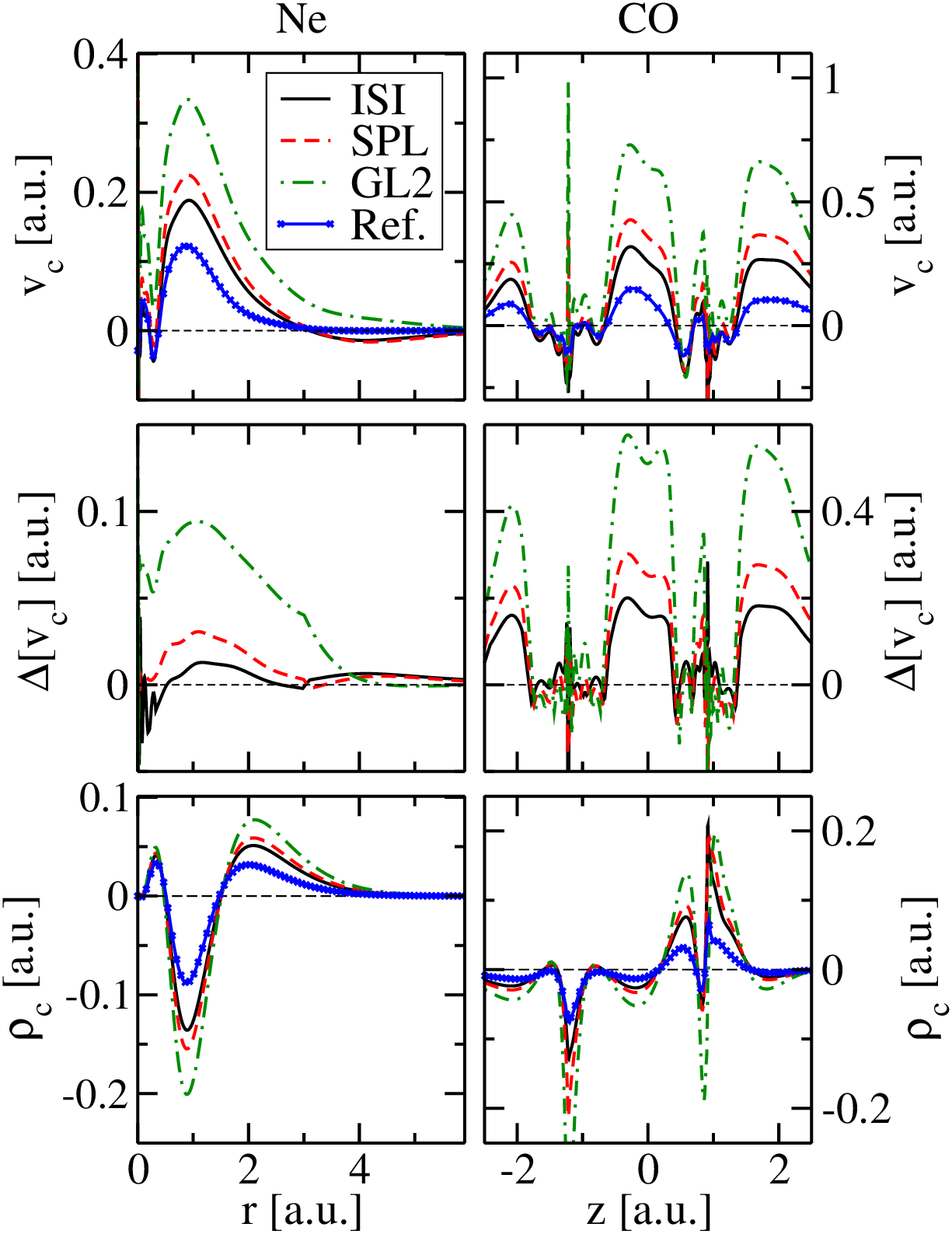}
\caption{Correlation potentials (top panels),
$\Delta[v_c]$ (middle) and correlation density (bottom) for the Neon atom (left)  and CO molecule (right) obtained using several ACM-SCF methods. Ref. means the CCSD(T) data using the method from Ref. \onlinecite{Wy2003}.}
\label{fig7}
\end{figure}

Then, we consider the correlation potentials for two typical systems, the Ne atom and the CO molecule.
In the top panels of Fig. \ref{fig7} we see that the ACMs provide a quite good description of the correlation potential for the two systems, improving significantly over GL2. 
Nevertheless, with respect to reference data there are still some limitations, e.g. a moderate overestimation of the correlation potential in valence regions.
This characteristic corresponds to an overestimation of shell oscillations in the SCF density, as indicated in the bottom panels of Fig. \ref{fig7},
where we report the correlation density $\rho_c$, i.e. the difference between the density obtained with a correlated method and its exchange-only version.

In the central panels of Fig. \ref{fig7}, we report the values
$\Delta[v_c(\R)]$, which is defined, in analogy to Eq. (\ref{eq:deltaee}) as
\begin{equation}
\Delta[v_c(\R)]=    |v_c@{\rm SCF}(\R)-v_c^{{\rm ref}}(\R)| -
    |v_c@{\rm EXX}(\R)-v_c^{{\rm ref}}(\R)|
    \label{eq:deltavc}
\end{equation}
These show, point-by-point whether or not the SCF procedure
improves the correlation potential with respect to EXX orbitals.
As we found for energies, the SCF correlation potentials are less accurate, but the error reduces with more accurate ACM functionals.
This feature is also evident for the correlation density, see bottom panels.
In this context, we should however also point out that the ACM-SCF density does not correspond to the exact linear response density
\cite{voora19,yang20,yufurche21}.

\begin{figure}[hbt]
\includegraphics[width=0.85\textwidth]{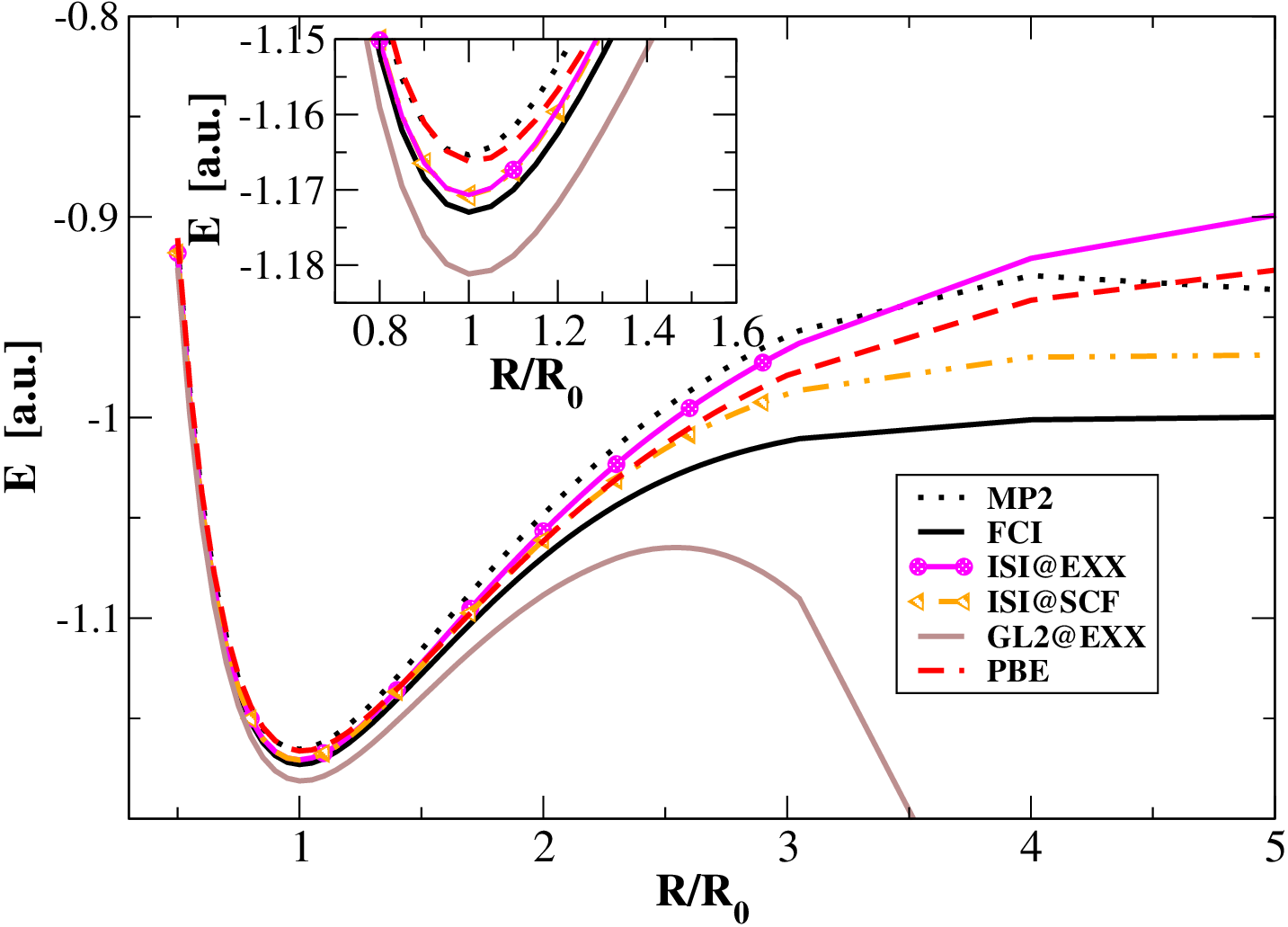}
\caption{The total energy of the H$_2$ molecule as it is stretched calculated with the various methods. The inset presents the same data around the equilibrium distance. }
\label{fig6}
\end{figure}

As a final case we consider in Fig. \ref{fig6} the potential energy surface for the dissociation of the H$_2$ molecule, in a restricted formalism \cite{cohen12}, which is one of the main DFT challenges \cite{cohen12,science21}, and was previously investigated in the ACM framework \cite{peach07,teale09,teale10}.  While both MP2 and GL2@EXX diverge at large distances, ISI@SCF nicely reproduces the exact FCI curve, much better than ISI@EXX, see also Ref. \onlinecite{fabianoisi16}. Thus the SCF procedure turns out to be quite important showing that, despite some limitations discussed above, it is crucial to include important correlation effects into the orbitals. For SPL (\del{not reported}\add{see Fig. S1 in the Supporting Information)} similar trends are found \add{the SPL@SCF curve for $R/R_0>2.5$ first increases and then decreases asymptotically, a behaviour which is clearly incorrect and depends on some drawbacks of the SPL functional to describe the limit for large distances, which is more influenced by the strong correlation.}.

 The limit for very large distances, well beyond $R/R_0>5$, is numerically tricky, but it can be \del{anyway} computed exactly using the hydrogen atom with fractional spins, H(1/2,1/2), i.e. with half spin up and half spin down \cite{cohen08}. For this system we have $E_{GL2}\rightarrow -\infty$ so that the ISI XC energy reduces to \cite{isi}
 \begin{equation}
     E_{xc}^{ISI} \rightarrow W_{\infty} + 2{W'}_{\infty} (1-\frac{1}{q}\ln(1+q))
 \end{equation}
 with $q=(E_x-W_{\infty})/{W'}_{\infty}$. The potential is thus a simple linear combination of  the EXX potential and the GGA potential  from  $W_{\infty}$ and ${W'}_{\infty}$.
 For the SPL approach, we have simply that 
 $E_{xc}^{SPL} \rightarrow W_{\infty}$ and thus the potential
 is just  $\delta W_{\infty}/\delta \rho({\bf r})$.
 
 The errors for different methods and orbitals are reported in Tab. \ref{tab:hh}.
 \begin{table}[hbt]
     \centering
     \begin{tabular}{rrrr}
     \hline
                 &     @EXACT & @SCF   & IDD\\  
                 \hline
          PBE    &      54.7 &   51.5  & 0.103 \\     
          EXX    &     196.1 &  178.6  & 0.260 \\ 
                 \hline
          SPL-PC  & \bf{-0.4}     &  - & - \\
          SPL-mPC & -109.8    & -114.9 & 0.125 \\
          SPL-hPC  & {\bf -21.0}    & {\bf  -21.4} & {\bf 0.024}\\ 
          \hline
         ISI-PC   & 27.4     &  - & - \\
         ISI-mPC &  90.2      &  83.7  & 0.151 \\
         ISI-hPC &  23.6    & {\bf 19.4} & {\bf 0.107}\\ 
         \hline
     \end{tabular}
     \caption{Total energy error for H(1/2,1/2) in kcal/mol for different methods and orbitals, \add{using a geometric series basis-set with 17 uncontracted Gaussian functions, $10^4$ as maximum exponent and 2.5 as geometric progression factor}.  The last column reports the integrated density difference error (IDD), i.e. $\int dr 4\pi r^2 |\rho(r)-\rho^{exact}(r)|$. \add{Note that self-consistent PC calculations do not converge. The best two ACM results are reported in boldface.}}
     \label{tab:hh}
 \end{table}
 At the exact density ($\rho(r)=\exp(-2r)/\pi$) SPL-PC gives an extremely accurate total energy \add{but the same method fails for the SCF calculation}. The SPL-mPC approach strongly underestimate the total energy, while  the SPL-hPC gives a much lower error\add{, both for the exact and the SCF densities}.
 At the ISI level all the energies are higher and the ISI-hPC\add{@SCF} is the most accurate approach. Note, however, that ISI-PC can be made exact with a proper choice of the parameters \cite{seidl00}. Note also that EXX fails for this system and PBE is also quite inaccurate.
 When SCF effects are considered, 
\textcolor{black}{When SCF effects are considered,  PBE, EXX, ISI-mPC, and ISI-hPC
     yield a slight improvement with respect to the case when the exact density is used. Because the integrated density difference (IDD) is not zero in all cases, this is a clear signature that all methods display some error compensation effect. Moreover, some methods give important convergence issues:} 
  the simple PC model does not converge, as explained above;
 the mPC model converges but the errors are very large, about twice the PBE ones.
 Instead, the ISI-hPC is very good \add{for both the considered densities}, having the best accuracy among all functionals and performing even better than all the functionals considered in Tab. 5 of Ref. \onlinecite{cohen12}.
Note
that the good accuracy of the ISI-hPC with respect to ISI-mPC is not related to the previously mentioned error cancellation between
an incorrect SCF density and an incorrect energy. In fact, \textcolor{black}{the 
IDD error is significantly smaller going from ISI-hPC  to ISI-mPC. Interestingly, the same arguments hold when comparing SPL-hPC to SPL-mPC, thus confirming the high quality of the hPC functional. Note that the almost vanishing IDD value for the SPL-hPC approach is a particular case, and all methods with $IDD\lesssim 0.1$ show a quite accurate density.}
The accuracy of the ISI-hPC@SCF approach for the H$_2$ dissociation limit is thus quite significant, considering that it uses full exact exchange and a combination of GL2 and a GGA functional without empirical parameters, in contrast to other approaches that use more complex constructions or extensive fitting on molecular data \cite{rev2021,science21}.
 
\section{Conclusions}
In this paper we have shown that it is possible to use ACM-based XC functionals in a full SCF procedure. This solves a long-standing issue in DFT as all the previous calculations with ACM functionals had been done in a post-SCF fashion using GGA or exact-exchange orbitals. This opens the way to new  applications and even basic studies in this context, removing the need for a post-SCF procedure and all the related sources of inaccuracy. Of course, despite the ACM-SCF procedure presented here is well defined, conceptually clean and fully capable of producing important results, is it fair to state that the whole method is not yet optimized and straightforward to apply especially because it is strictly related to the OEP approach used for the treatment of the GL2 component, which requires itself some expertise to be handled. Nevertheless, several tricks and improvements can be used to make the OEP calculations simpler and more reliable \cite{yang08}, thus various upgrades can be easily seen from the practical point of for the SCF-ACM method. Anyway, these are left for future works as in this paper we wanted to focus only on the core of problem without adding too many technical details.  

Having been able to perform SCF ACM calculations on various systems we could perform a thorough assessment of the \del{functionanls} \add{functionals}, finding important results.
For strongly-correlated systems, such as the harmonium atom and the hydrogen molecule at the dissociation limit, the ACM SCF calculations yield very accurate results taking advantage of the incorporated strong-correlation limit and also thanks to the novel hPC functional for $W_{\infty}$ and ${W'}_{\infty}$ that proved to be very accurate for these cases.
For molecular systems, we found that the overall accuracy using SCF orbitals depends on the quality of the underlying ACM, in line with the Refs. \onlinecite{bleiz13} and \onlinecite{bleiz15}. In any case the ISI-hPC  yields already quite correct SCF potentials and total energies: \add{nevertheless its accuracy need to be further verified for reactions and atomization energies}

Thus, we can finally conclude that, despite some limitations, the overall accuracy of the ISI functional (and partially also of the SPL one), when the full SCF solution is taken into account, is overall satisfactory, especially considering that: 
i) it does not employ any parameter obtained from molecular systems, ii) the approach is within a pure KS formalism with a local potential.
These results and the availability of a working SCF procedure for general ACM formulas now open to the application and testing on other systems beyond the simple ones considered in this work. Moreover, it paves the path towards the development of more accurate ACM functional forms
(see e.g. Ref. \onlinecite{spl2}) as well as to further development of $W_{\infty}$ and ${W'}_{\infty}$ approximations, with improved accuracy for molecular systems.

\section*{Supporting Information}
Details on the basis set, dissociation curve of $H_2$ with the SPL functional, further results for the Hooke's atom.

\section*{Acknowledgements}
S.\'S. thanks the Polish National Science Center for the partial financial support under Grant No. 2020/37/B/ST4/02713 whereas E.F. and F.D.S. thanks for financial support the CANALETTO project (No. PPN/BIL/2018/2/00004, PO19MO06). PG-G was funded by the Netherlands Organisation for Scientific Research (NWO) under Vici grant 724.017.001.

\appendix
\section{Interpolation Formulas} \label{app:formulas}
In the following we report the ISI and SPL interpolation formulas. \\

\noindent{\bf  Interaction Strength Interpolation (ISI) formula} \cite{seidl00}\\
\begin{equation}
W_\lambda^\ISI  = W_\infty + \frac{X }{\sqrt{1+\lambda Y }+Z }\ ,
\end{equation}
with
\begin{eqnarray}\label{Y}
&&X=\frac{xy^2}{z^2}\; ,\; Y=\frac{x^2y^2}{z^4}\; , \; Z=\frac{xy^2}{z^3}-1\ ;\\
&& x=-2 W_0' ,\; y=W_\infty'\; , \; z=W_0-W_\infty\ ,
\end{eqnarray}
which yields for the exchange-correlation energy:
\begin{equation}
E_{xc}^\ISI = W_\infty + \frac{2X}{Y}\left[\sqrt{1+Y}-1-Z\ln\left(\frac{\sqrt{1+Y}+Z}{1+Z}\right)\right]\ .
\end{equation}
\noindent{\bf Seidl-Perdew-Levy (SPL) formula} \cite{SeiPerLev-PRA-99}\\
\begin{equation}\label{spl_eq}
W_\lambda^\SPL  = W_\infty  +\frac{W_0 -W_\infty }{\sqrt{1+2\lambda \chi }}\ ,
\end{equation}
with
\begin{equation}
\chi = \frac{W_0'}{W_\infty-W_0}\ .
\end{equation}
The SPL XC functional reads
\begin{equation}
E_{xc}^\SPL = \left(W_0-W_\infty\right)\left[\frac{\sqrt{1+2\chi}-1-\chi}{\chi}\right] + W_0\ .
\end{equation}
Notice that this functional does not make use of the information from $W_\infty'$.\\

\bibliography{main}

\end{document}